\documentclass[usenatbib]{mnras}
\usepackage[T1]{fontenc}
\usepackage{ae,aecompl}

\usepackage{graphicx}	
\usepackage{amsmath}	
\usepackage{amssymb}	
\usepackage{tikz}

\newcommand{\jpb}{{J. Phys. B: At. Mol. Opt. Phys.}}
\newcommand{\rpp}{{Rep. Prog. Phys.}}
\newcommand{\rmp}{{Rev. Mod. Phys.}}

\title[Photoionization of Zn II ions]{Single photoionization of the Zn II ion in  
                                                        the photon energy range 17.5 to 90.0 eV: experiment and theory}
\author[G. Hinojosa et al]{G. Hinojosa$^{1}$\thanks{E-mail:hinojosa@icf.unam.mx}, V. T. Davis$^{2}$, 
                    A. M. Covington$^{2}$, J. S. Thompson$^{2}$, 
\newauthor{A. L. D. Kilcoyne$^{3}$, A. Antill\'{o}n$^{1}$,  E. M. Hern\'{a}ndez$^{1,7}$,  
                    D. Calabrese$^{4}$,}  
\newauthor{A. Morales-Mori$^{1}$, A. M. Ju\'{a}rez$^{1}$, O. Windelius$^{5,6}$ and 
                    B. M. McLaughlin$^{8,9}$\thanks{E-mail:bmclaughlin899@btinternet.com}
}
\\
\\
$^{1}$Universidad Nacional Aut\'{o}noma de M\'{e}xico, Instituto de Ciencias F\'{i}sicas, AP 48-3, Cuernavaca 62251, M\'{e}xico\\
$^{2}$Department of Physics, University of Nevada, Reno, NV 89557-0220, USA\\
$^{3}$The Advanced Light Source, Lawrence Berkeley National Laboratory, Berkeley, California 94720, USA\\
$^{4}$Department of Physics, Sierra College, Rocklin, CA 95677, USA\\
$^{5}$Department of Applied Physics, Chalmers University of Technology, S-41296 G\"{o}teborg, Sweden\\
$^{6}$Department of Physics, University of Gothenburg, SE-412 96, G\"{o}teborg, Sweden\\
$^{7}$Universidad Aut\'{o}noma del Estado de Morelos, Avenida Universidad 1001, Cuernavaca 62209, M\'{e}xico\\
$^{8}$Centre for Theoretical Atomic and Molecular Physics (CTAMOP),
           School of Mathematics and Physics,\\
	 Queen's University Belfast, Belfast BT7 1NN, UK\\
$^{9}$Institute for Theoretical Atomic and Molecular Physics (ITAMP)\\
	Harvard Smithsonian Center for Astrophysics, 
	 MS-14, Cambridge, MA 02138, USA\\
		}
\date{Accepted \today. Received \today; in original form May 22, 2017}

\pubyear{2017}

\begin{document}

\label{firstpage}
\pagerange{\pageref{firstpage}--\pageref{lastpage}}

\maketitle

\begin{abstract}
{
Measurements of the single photoionization cross section of Cu-like Zn$^+$ ions are reported in the energy (wavelength) 
range 17.5 eV (709 \AA) to 90 eV (138 \AA). The measurements on this {\it trans}-Fe element 
were performed at the Advanced Light Source synchrotron radiation facility 
in Berkeley, California  at a photon energy resolution of 17 meV 
using the photon-ion merged-beams end-station. 
Below 30 eV the spectrum is dominated by excitation autoionizing resonance states. 
The experimental results are compared with large-scale photoionization cross-section calculations 
performed using a Dirac Coulomb $R$-matrix  approximation. 
Comparison are made with previous experimental studies, resonance states are identified 
and contributions from metastable states of Zn$^+$ determined.
}
\end{abstract}



\section{Introduction}
About half of the heavy elements (Z $\ge$30) in the Universe 
were formed during the Asymptotic Giant Branch (AGB) 
phase through slow  neutron-capture (\emph{n}-capture) nucleosynthesis  (the  \emph{s}-process).  
In  the  intershell  between  the  H-  and  He-burning  shells, Fe-peak nuclei experience 
neutron captures interlaced with $\beta$-decays, which transform them into heavier elements.
For example, in spite of its cosmic rarity, atomic selenium has been detected in the
spectra of stars \citep{Roederer2012,Roederer2014}
 and of astrophysical nebulae \citep{Sterling2008,Sterling2009,Garcia2015,Garcia2016}.  
The chemical composition of these objects illuminate details of stellar 
nucleosynthesis and the chemical evolution of galaxies.  
To interpret such astrophysical spectra requires three types of fundamental 
parameters: i) the energy of internal states (or transition frequencies), ii) transition
probabilities (or Einstein A-coefficients), and iii) collisional 
excitation rate coefficients (or collision
strengths in the case of electron impact). 
While these parameters have been studied over the
past century for many atomic and molecular species, 
the necessary information is far from complete or is of insufficient
reliability. Extensive databases do exist for the first two items (e.g., NIST),
referred to as spectroscopic data; the situation for the last item is far less satisfactory.

However, atomic data, such as PI cross sections, are unknown 
for the vast majority of \textit{trans}-Fe, neutron-capture element ions.  
We note that the \emph{n}-capture elements, Cd, Ge, and Rb, 
were recently detected in planetary nebulae \citep{Sterling2016}.  
Measurements for the photoionization cross section 
on the \textit{trans-Fe} element Se, in various ionized stages
\citep{Sterling2011,Sterling2011a,Sterling2011b,Esteves2011,Esteves12,Macaluso2015} 
provided critically needed data to astrophysical modelers 
\citep{Sterling2011c,Sterling2015,Garcia2016}, and was benchmarked
against results from a recently developed suite of fully relativistic parallel Dirac 
Atomic {\it R}-matrix (DARC) codes 
\citep{darc,McLaughlin2012,Ballance2012,McLaughlin2015a,McLaughlin2016,McLaughlin2017} 
with excellent agreement being achieved.

Studies of Zn I abundance  in metal-poor stars are used to 
infer the histories of galactic chemical evolution. 
Unfortunately, a major source of 
uncertainty in all these studies is a lack of accurate knowledge 
of atomic parameters, such as transition wavelengths, 
cross sections for photoionization (PI), electron-impact excitation (EIE),
dielectronic recombination (DR), and line-strengths 
\citep{Sneden1991,Mishenina2002, Roederer2011,Roederer2010,Frebel2014}.

Photoionization of atomic ions is an important process in determining
the ionization balance and the abundances of elements
in photoionized astrophysical nebulae. It has recently become
possible to detect neutron \emph{n}-capture elements 
(Se, Cd, Ga, Ge, Rb, Kr, Br, Xe, Ba and Pb) in a large number
of ionized nebulae \citep{Sterling2007,Sharpee2007,Sterling2008}. 
These elements are produced by slow or rapid  \emph{n}-capture nucleosynthesis. 
Measuring the abundances of these elements helps to reveal their dominant production
sites in the Universe, as well as details of stellar structure,
mixing and nucleosynthesis \citep{Sharpee2007,Lan2001,Sterling2016}. 
These astrophysical observations are the motivation to determine the EIE, EII, PI, and
recombination properties of \emph{n}-capture elements 
\citep{Card1993,Smith1990,Wall1997,Busso1999,Lan2001,Trav2004,Herwig2005,Sneden2006}.

The validity of various  theories of galactic evolution, stellar nucleosynthesis, 
the interplay of gravity and chemistry 
in various stellar bodies all depend on accurate measurements of metals 
within different types of astrophysical objects such  as large stars, nebulae and globular clusters 
\citep{Sneden1991,Chayer2005,Mishenina2002,Roederer2011,
Jose2011,Hinkel2014,Nissen2007,Nissen2004,Chen2004,Hollek2011,Kobayashi2009,Djenize2003}.
The oldest parts of the Milky Way galaxy are the metal-poor stars in the thick disk. 
The origin and evolution of the chemical and dynamical structures of the thick disk remain unclear. 
Chemical abundances in stars of the thick disk, and in the inner 
and outer halo stars are important because the ratios of 
heavy elements to Fe in these stars can reveal different 
formation mechanisms. [Zn/Fe] ratios in outer halo stars can 
be used to distinguish formation timescales of these stars from those 
of stars in the disk \citep{Nissen2004,Chen2004,Tumlinson2006}.

The relevance that atomic parameters have in astrophysics and 
in our understanding of the chemical 
evolution and the chemical assembly of progenitor nebulae is 
derived from measurements of elemental abundances of zinc. 
In these nebulae, zinc happens to be a better indicator of 
Fe and Fe-group abundances,  as it does not condense into dust 
as easily as iron. Some nebulae do not show the abundance of 
Zn that models predict, indicating that reliable data on this 
species are of critical importance in the study of these systems  \citep{Karakas2009}.

Experimental studies on the photoionization (PI) of ions 
from intermediate-sized atoms provide a 
wealth of spectroscopic data and cross section values. 
PI cross section data on {\it trans}-Fe ions can be used to advance 
theoretical models in light of details that substantial photon 
flux and high energy resolution have the power to reveal. 
To our knowledge, the only data available on the PI of the {\it trans}-Fe element, Zn II, 
is the pioneering experimental work by \citet{Peart1987}, using a merged-beams technique
 by combining synchrotron radiation (SR) with a beam of Zn$^{+}$ 
 at the Daresbury radiation facility. 
\citet{Peart1987} succeeded in observing several resonances 
and  measured the photoionization cross section 
with moderate photon energy resolution, at the Daresbury 
beam line \citep{Lyon1986,Lyon1987,Peart1987}, which 
had an energy resolution ($\Delta \lambda$)  ranging from 0.2 -- 1 \AA, 
or approximately 20 meV -- 4 meV \citep{Lyon1986}.  
Since the photon flux for measurements on Zn II ions \citep{Peart1987} was low, 
a direct measurement of  the background photoionization 
cross section was not possible nor was identification of the peaks in the spectrum. 
The previous work of \citet{Peart1987} however may be used 
for comparison  purposes and verification 
of the data presented here.

Previous studies on the PI of Zn I and species iso-electronic to Zn II,  relevant to the present 
work have been performed by \citet{Harrison1969}, who studied PI of the neutral Zn atom,
using UV discharge sources.  \citet{Mueller1986} studied PI of the iso-electronic neutral Cu atom. 
From a theoretical point of view, Zn I is the first transition metal with a 
closed $d$-shell with possible $np$ resonances \citep{Stener1997}. 
Apart from its fundamental importance, in astrophysics, 
photoionization of Zn$^+$  (Zn II) is of great practical interest. 
 Recently, \citet{Ganeev2016} demonstrated that 
autoionizing states of Zn II and Zn III have 
the ability to enhance harmonics in laser-produced plasmas. 
We note the K$_{\alpha}$ X-ray spectrum of the Cu atom 
has also been measured recently by \citet{Cline2017}.
Autoionizing states are, in general, excited states with energies 
that are not commonly studied in the literature.  In the present work, 
energies for several of these intermediate autoionizing resonant states 
have been measured and identified. 

The prime motivation for the present study of this {\it trans}-Fe element, Zn II,
 is to provide benchmark PI  cross section data for applications in astrophysics.
High-resolution measurements of the photoionization cross section of Zn$^{+}$  
were performed at the ALS synchrotron radiation facility in Berkeley, California, 
over the photon energy range 17.4 -- 90 eV at a resolution of 17 meV FWHM.  
Several highly excited states have been identified in the energy (wavelength) 
range 20 eV (620 \AA) to 90 eV (138 \AA). The high resolution ALS 
photoionization cross sections are used to benchmark 
large-scale \textsc{darc} calculations in this same energy interval.
A comparison of the \textsc{darc} PI cross sections 
with previous experimental studies \citep{Peart1987}
and the present ALS work indicate excellent agreement, 
providing further confidence in the data for 
various astrophysical applications.

\section{Experiment}
The present experimental technique and apparatus have been described in 
detail previously by \citet{Covington2002}. 
Recent improvements to the technique have been discussed by M\"{u}ller and 
collaborators \citep{Mueller2014,Mueller2015}. 
Here, we present a brief description with details relevant to the present measurements.  
The experimental method is known as the merged-beams technique \citep{peart1973,Lyon1986,Lyon1987}.  
The method consists of overlapping trajectories of two beams; in this case, 
a photon beam from the Advanced Light Source (ALS) 
synchrotron at the Lawrence Berkeley National Laboratory 
and a counter-propagating ion beam of Zn$^{+}$ ions. 
The experiment was designed to count the resulting 
Zn$^{2+}$ ions and measure the relevant parameters of 
the ion and photon beams as well as their spatial overlap.

The photon beam was generated by a 10-cm-period undulator located in the synchrotron ring. 
The synchrotron was operated under an almost constant electron current of 0.5 A at 1.9 GeV. 
The resulting photon beam had a maximum width of 1.5 mm and a divergence less than 0.06$^o$.  
In the grazing incidence mode, the photon beam was directed onto a spherical grating with controls 
that allowed changes to the photon beam energy to be made as desired. 

We used a side-branch gas cell with either He or Kr gases to calibrate the photon energy in the 
energy range from 18.601 eV to 92.437 eV \citep{Domke1996,King1977}. 
A polynomial fit to reference energy values 
was used to calibrate the energy scale (along with the Doppler shift correction due to the  
counter-propagating photon and ion beams). As a result of the calibration, our data is in agreement with 
the values reported by NIST \citep{NIST2016} to 3 significant figures. The uncertainty associated with this 
procedure is estimated to be $\pm$10 meV. The Zn$^{+}$  ion beam was produced with an 
all-permanent magnet ECR ion source by evaporating zinc in an oven that was inserted into 
the ion source chamber, with argon used as a support or buffer gas.   A 60$^o$ - sector analyzing 
magnet was used to separate (as a function of its momentum-to-charge ratio) the Zn$^{+}$  
cations which had a kinetic energy of 6 keV.   A cylindrical Einzel lens, two sets of steering plates, 
and a set of slits were used to focus and collimate the ion beam. A set of electrostatic 90$^o$ spherical-sector 
plates was used to merge the ion beam with the photon beam. 

The ions then entered a voltage-biased  cylindrical interaction region (IR), 
Zn$^{2+}$  photo-ions produced inside it had a different energy 
as compared to those produced outside, thereby energy-labeling the ions generated within the specific 
length of the IR. Energy-tagged Zn$^{2+}$ ions were subsequently separated from the primary 
Zn$^{+}$ ion beam by a 45$^o$ dipole analyzing magnet. The experiment was carried out in two different 
operational modes. The purpose of the first operational mode was to obtain precise knowledge of the 
interaction length and beam overlap in order to measure the cross section. In this operational mode, 
the overlap of the two beams within the IR was carefully measured at discrete photon energies. 

An absolute-calibrated photo diode was used while operating in this mode. In the second mode, the 
photo-ion signal was maximized without monitoring the actual overlap of the beams within the IR. 
While running in this mode, the IR voltage was also maintained at all times and the beams overlap 
was monitored at only some photon energies between successive energy range scans, resulting 
in a photo-ion yield spectrum that was later normalized to the absolute measurements. The purpose 
of measuring spectroscopic energy-scans with the IR voltage on was to measure the absolute cross 
sections by slightly tuning back the overlap at the pre-established photon energies. Integration 
times of 200s were used to account for the stabilization time of the photo-diode current reading.  

The Zn$^{+}$ ion beam was measured by an extended Faraday cup. The resulting Zn$^{2+}$ 
photo-ions were counted of the dipole analyzing magnet whose magnetic field was adjusted 
so that only product photo-ions generated inside the interaction region were collected. In addition, 
by using a chopper wheel in the photon beam, the Zn$^{2+}$ signal was background-subtracted 
to remove those ions that were produced by collisions with residual gas in the IR. To derive absolute 
PI cross sections, two-dimensional beam profiles of the ion beam $I^{+}(x,y)$ and the photon beam 
$I(x,y)$ were measured and used to calculate the interaction volume of the two beams. 
Three beam translating-slit profilers were used to sample the form factor $F(z)$ according to 
\begin{equation}
 F(z)= \frac{ \int_{}^{} {\int_{}^{} I^{+} (x,y) I^{\gamma}(x,y) dx dy}} 
                                      { \int_{}^{}  I^{+} (x,y) dx dy   \int  I^{\gamma}(x,y) dxdy } 
\label{factor}
\end{equation}
where $z$ is the reference axis assigned to the propagation direction of the ion beam. 
$F(z)$ was measured at three positions, at the entrance, in the centre and at the exit of the IR. 
Measured values of these positions were labeled as $i$ =1, 2, 3, respectively. With these three 
values of $F_i (z)$, $F(z)$ was interpolated within the IR length and integrated over $z$ to 
compute the spatial overlap of the photon and ion beams along the common IR path.

\begin{table}
\centering
\caption{Zn$^{+}$ absolute single-photoionization valence shell cross section made 
               at the ALS with a 17 $\pm$ 3 meV photon energy resolution in the photon energy 
               range 25 - 90 eV. \label{absolute}} 
\begin{tabular}{cc}
\hline
Energy  (eV) &  $\sigma$ (Mb)  	\\
\hline
  25.0			&\; 1.14	$\pm$  0.23\\
  29.8			&\; 7.83	$\pm$ 1.60\\
  35.0			&\; 9.80	$\pm$ 1.80\\
  45.0			&\; 9.99	$\pm$ 2.00\\
  60.0			&11.20	$\pm$ 2.24\\
  78.0			&\; 9.73	$\pm$ 1.95\\
  89.5			&10.60	$\pm$ 2.14\\
\hline
\end{tabular}
\end{table}

Absolute cross section measurements were performed at specific 
photon energies (open circles (magenta) in Fig.\ref{fig1} (a) and (b))  
where no resonant structure was present in the spectrum and are tabulated in Table \ref{absolute}. 
The single-photoionization cross section for Zn$^{+}$ was derived from the expression
\begin{equation}
\sigma = \frac{R q e^{2} \nu_i  \epsilon}{I^{+} I^{\gamma} \int_{}^{} F(z) dz},
\label{crossx}
\end{equation}
where $R$ is the photo-ion count rate, $q$=1 is the 
charge state of Zn$^{+}$, $e$=1.6$\times$ 10$^{-19}$ C,
$\nu_i$ is  the ion beam velocity in cm/s,  $\epsilon$ 
is the responsivity of the photo-diode (electrons per photon), 
$I^{+}$ is the ion beam current (A), and $ I^{\gamma}$  is the photo-diode current (A). 
In this experiment, the photo-ion counting efficiency was practically 100\%.

The largest sources of systematic error originate from the beam-overlap integral, 
the beam profile measurements, and the photo-diode responsivity function. Other contributions 
to the total systematic error are listed in \citet{Covington2002}. All sources of systematic error 
combine to yield a total uncertainty of 20\%, estimated at the 90\% confidence level. 

The spectra shown in Fig.\ref{fig1} and in panel (a) of Fig. \ref{fig2} were measured 
at 5 to 10 eV-wide photon energy intervals. 
Each interval overlapped its neighbouring intervals by 1.0 eV.  
All the individual parts of the spectra 
were later combined by joining adjacent regions 
to produce the entire measured spectrum.  
An estimation of the photon energy uncertainty caused 
by this gluing procedure was not greater than $\pm$8 meV. 

The overall energy uncertainty propagated by both the 
gas cell energy calibration and the data reduction 
procedure is $\pm$13 meV. The photo-ion yield spectra 
were normalized by using the measurements at the 
discrete photon energies of the absolute photoionization cross section. 

\begin{table}
\centering
\caption{The correction factor $f_c$ applied to the present measured ALS cross section data
         due to the presence of high-order radiation effects in the photon beam. In our work,
         we interpolated the values from \citet{Mueller2015} and included values at 30 eV and 35 eV.}
\label{correct} 
\begin{tabular}{cc}
\hline
Photon energy  (eV) & Correction factor  $f_c$ 	\\
\hline
  16			& 1.90$^a$	\\
  20			& 1.40$^a$	\\
  25			& 1.17$^a$	\\
  30			& 1.07$^b$	\\
  35			& 1.00$^b$	\\
\hline
\end{tabular}
\begin{flushleft}
$^a$Values determined by \citet{Mueller2015}.\\
$^b$Present values.\\
\end{flushleft}
\end{table}
It is important to point out that corrections in the cross section caused by higher-order radiation in the photon beam 
(mainly second- and third-order radiation in the lower energy regime) can be significant in this photon beam line. 
The correction to the cross section at 20 eV was estimated to be almost 40\% by \citet{Mueller2015} 
who derived a correction function $f_c$ for this photon beam-line. The error of this correction is 50\% of the 
difference between the uncorrected and the higher-order-corrected cross section. 

Table \ref{correct} gives the correction factor $f_c$ to the cross section due to high-order radiation effects 
in the photon beam for the present measurements used in our work. Instead of directly applying 
the correction function of \citet{Mueller2015}, we used an approximation by interpolating 
the values of $f_c$ from \citet{Mueller2015} listed in Table \ref{correct}, 
and included values at 30 eV and 35 eV. 
From this approximation, we found that at 17.4 eV the total uncertainty is 
33\% and, at 19.64 eV the total uncertainty is already below 20\%. Hence for the present experiment 
we quote a total error of 33\%  below 19.64 eV and 20\% for higher energies in the spectrum.

\section{Theory}
\subsection{Atomic Structure}
The \textsc{grasp} code \citep{Dyall1989,Grant2006,Grant2007} was used to generate 
the target wave functions employed in our collisions work.  All orbitals were 
physical up to $n$=3, 4$s$ and 4$p$. We began by doing an 
extended averaged level (EAL) calculation for the $n$ = 3 orbitals .
All EAL calculations were performed on the lowest 24 fine-structure levels of the residual 
Zn III ion,  in order to generate target wavefunctions for our photoionization studies. 
In our work we retained all the 355 - levels originating from 
one and two-electron promotions from the $n$=3 levels into the orbital space of this ion. 
All  355 levels from the sixteen configurations 
were included in the \textsc{darc} close-coupling calculation, namely:
$3s^23p^63d^{10}$, 
$3s^23p^63d^94s$,   $3s^23p^63d^94p$,
 $3s^23p^53d^{10}4s$, $3s^23p^53d^{10}4p$,
$3s3p^63d^{10}4s$,     $3s3p^63d^{10}4p$,
$3s^23p^63d^84s^2$,   $3s^23p^63d^84p^2$,  $3s^23p^63d^84s4p$, 
$3s^23p^43d^{10}4s^2$, $3s^23p^43d^{10}4p^2$, $3s^23p^43d^{10}4s4p$, 
$3p^63d^{10}4s^2$,        $3p^63d^{10}4p^2$ and $3p^63d^{10}4s4p$. 

Table \ref{Levels} gives a sample of the theoretical energy levels from the 355-level 
\textsc{grasp} calculations for the lowest 17 levels of the residual Zn$^{2+}$ ion, compared to the 
values available from the NIST tabulations \citep{NIST2016}.   The average percentage 
difference of our theoretical energy levels compared with the NIST values is approximately 12\%.  
For the $3d^{10}\; {\rm ^1S_0} \rightarrow 3d^94p\; {\rm ^3P^o_1}$ transition we obtained 
a value of 4.0$\times$10$^{-3}$ for the oscillator strength ($f$-value). 
This is to be compared with the $f$-value of 
5.0$\times$10$^{-3}$, for the same transition 
from the \textsc{mcdf} work of \citet{Yu2007}.
From our \textsc{grasp} calculations, we obtained values 
of 1.3 $ns$ and 1.8 $ns$ for the radiative lifetimes for 
the 1581~\AA ~and 1673~ \AA~ lines, respectively, the 
$3d^94p\;{\rm ^3F^o_4} \rightarrow 3d^94s\;{\rm ^3D_3}$ 
and $3d^94p\;{\rm ^3P^o_2} \rightarrow 3d^94s\;{\rm ^3D_3}$ 
transitions in Zn III.  Our values are in 
respectable agreement with the experimental results of    
1.1 $\pm$ 0.3 $ns$ and 1.2 $\pm$ 0.3 $ns$  from 
the previous work of \citet{Andersen1977}.

Photoionization cross sections were then performed on the Zn II ion
for the $3d^{10}4s\; {\rm ^2S}_{1/2}$ ground state 
and  the $3d^{9}4s^{2}\; {\rm ^2D}_{3/2,5/2}$ metastable levels 
 using the \textsc{darc} codes, with these Zn III residual ion  target wavefunctions. 

\subsection{Photoionization calculations}
For comparison with high-resolution measurements 
made at the ALS, state-of-the-art theoretical methods 
with highly correlated wavefunctions that 
include relativistic effects are used.
 The scattering calculations were performed for photoionization cross sections using 
the above large-scale configuration interaction (CI) target wavefunctions  as input
to the parallel \textsc{darc} suite of $R$-matrix codes.

The General Relativistic Atomic Structure Package (\textsc{grasp}) \citep{Dyall1989,Grant2007} 
formulates and diagonalises a Dirac-Coulomb Hamiltonian \citep{Grant2007} to produce the relativistic orbitals 
for input to the Dirac Atomic $R$-matrix Code (\textsc{darc}) 
\citep{Chang1975,Norrington1981,Grant2007,Burke2011} to obtain the photoionization (PI) cross sections.  
These \textsc{darc} codes are currently running efficiently on a variety of world-wide 
parallel High Performance Computer (\textsc{hpc})  architectures
 \citep{darc,McLaughlin2015a,McLaughlin2015b,McLaughlin2016,McLaughlin2017}. 

Fifteen continuum orbitals were used in our scattering calculations.
A boundary radius of 9.61 a$_0$ was necessary to accommodate the 
diffuse $n$ = 4  bound state orbitals of the residual Zn III  ion. To fully resolve
the resonance features present in the spectrum, an 
energy  grid of 2.5 $\times$ 10$^{-7}$ ${\cal{Z}}^2\;Ry$ (13.6  $\mu$eV), 
where $\cal{Z}$ = 2,  was utilized in our collision work.  

Photoionization cross section calculations with this 355-level model 
 were then performed with this fine energy mesh for the $3d^{10}4s\; {\rm ^2S}_{1/2}$ ground state 
and the  $3d^{9}4s^{2}\; {\rm ^2D}_{5/2,3/2}$ metastable levels 
of this ion, over the photon energy range similar to experimental studies.  
This insured that all the fine resonance features were fully resolved in the 
respective PI cross sections.  

For the ${\rm ^2S}_{1/2}$ level we require the bound-free dipole 
 matrices, J$^{\pi}$ = 1/2$^e$, \ $\rightarrow$ \ J$^{{\prime}\pi^{\prime}}$ = 1/2$^o$ and 3/2$^o$. 
In the case of the ${\rm ^2D}_{5/2,3/2}$ metastable levels we required the 
J$^{\pi}$ = 3/2$^e$, \ $\rightarrow$ \ J$^{{\prime}\pi^{\prime}}$ = 1/2$^o$ ,  3/2$^o$,  5/2$^o$ 
and the J$^{\pi}$ = 5/2$^e$, \ $\rightarrow$ \ J$^{{\prime}\pi^{\prime}}$ = 3/2$^o$ ,  5/2$^o$,  7/2$^o$ 
bound-free dipole matrices.   The $jj$-coupled Hamiltonian diagonal matrices  were adjusted so that 
the theoretical term energies matched the recommended experimental values of the 
NIST tabulations  \citep{NIST2016}. We note that this energy adjustment  ensures 
better positioning of resonances relative 
to all thresholds included in the present calculations.

\begin{table}
\centering
\caption{Zn$^{2+}$ energy levels in Rydbergs (Ry) from the large-scale \textsc{grasp} calculations compared with 
               the available tabulations from the NIST database \citep{NIST2016}.  A sample of the lowest 17 levels 
              for the residual Zn$^{2+}$ ion from the 355-level \textsc{grasp} calculations are shown
              compared to experiment. The percentage difference $\Delta (\%)$ of individual energy 
              levels is given for completeness. \label{Levels}} 
\begin{tabular}{cccccc}
\hline
Level	 	&STATE		& TERM			&  NIST		&GRASP			& $\Delta (\%)^a$  	\\
		&			&				&  (Ry)		& (Ry)				&			\\								
\hline
1		&$3d^{10}$		& $\rm ^1S_0$		& 0.000000		&0.000000			& 0.0			\\
\\
2		&$3d^94s$		&$\rm^3D_3$		& 0.711666   	&0.820402			&15.3			\\
3		&$3d^94s$		&$\rm^3D_2$		& 0.722394 		&0.834832			&15.6			\\
4		&$3d^94s$		&$\rm^3D_1$		& 0.736761     	&0.851328			&15.6			\\
\\
5		&$3d^94s$		&$\rm^1D_2$		& 0.760911 		&0.893474			&17.4			\\
\\
6		&$3d^94p$		&$\rm^3P^o_2$		&1.256331      	&1.361426			&8.4			\\
7		&$3d^94p$		&$\rm^3P^o_1$		&1.276421 	   	&1.385355			&8.5			\\
8		&$3d^94p$		&$\rm^3P^o_0$		&1.288463   		&1.399012			&8.6			\\
\\
9		&$3d^94p$		&$\rm^3F^o_3$		&1.281741 		&1.410420			&10.0			\\
10		&$3d^94p$		&$\rm^3F^o_4$		&1.287866    	&1.411489			&9.6			\\
11		&$3d^94p$		&$\rm^3F^o_2$		&1.298403  		&1.427982			&10.0			\\
\\
12		&$3d^94p$		&$\rm^1F^o_3$		&1.316792    	&1.457057			&10.7			\\
\\
13		&$3d^94p$		&$\rm^1D^o_2$		& 1.323560  		&1.468436			&10.9			\\
\\
14		&$3d^94p$		&$\rm^3D^o_3$		& 1.330146 		& 1.472693			&10.7			\\
15		&$3d^94p$		&$\rm^3D^o_1$		& 1.344768   	& 1.491643			&11.0			\\
16		&$3d^94p$		&$\rm^3D^o_2$		&  1.347960    	& 1.496767			&11.0			\\
\\
17		&$3d^94p$		&$\rm^1P^o_1$		&  1.344106		&1.527372			&13.6			\\
\hline
\end{tabular}
\begin{flushleft}
$^a$Average $\Delta (\%)$ of the energy levels with experiment is $\approx$ 12\%.
\end{flushleft}
\end{table}

\subsection{Resonances}
The energy levels available from the NIST tabulations  \citep{NIST2016} 
were used as a helpful guide for the present assignments.
The resonance series identification can be made from Rydberg's
 formula \citep{rydberg1,rydberg2,rydberg3,eisberg1985} :
\begin{equation}
	\epsilon_n= \epsilon_{\infty} - \frac{{\cal{Z}}^2}{\nu^2}
\end{equation}
where in Rydbergs $\epsilon_n$ is the transition 
energy, $\epsilon_{\infty}$ is the
ionization potential of the excited 
electron to the corresponding final state ($n$ = $\infty$),
 i.e. the resonance series limit with 
 $n$ being the principal quantum number.
The relationship between the principal quantum number
$n$, the effective quantum number $\nu$,  and the quantum 
defect $\mu$ for an ion of effective charge $\cal{Z}$ is given by 
 $\nu$ = $n$ - $\mu$  \citep{Shore1967,Seaton1983}. 
 Converting all quantities to eV, we 
 can represent the Rydberg series as
\begin{equation}
	E_n= E_{\infty} - \frac{{\cal{Z}}^2 {\cal R}}{(n - \mu)^2.}
\label{rydberg}
\end{equation}
Here, $E_n$ is the resonance energy, $E_{\infty}$ the resonance series
limit, $\cal{Z}$  is the charge of the core (in this case $\cal{Z}$ = 2), $\mu$
is the quantum defect 
($\mu_{ns}$ ~$>$ ~$\mu_{np}$ ~$>$~ $\mu_{nd}$~ $>$ ~$\mu_{nf}$, $\dots$), 
 being zero for a pure hydrogenic
state.  For a hydrogenic system this can be written as
\begin{equation}
	E^{\rm H}_n= E_{\infty} - \frac{{\cal{Z}}^2 {\cal R}}{n^2},
\end{equation}
where the Rydberg constant $\cal R$ is $R_{\infty}$  = 13.605698 eV.

The multi-channel {\it R}-matrix eigenphase derivative
(QB) technique, which is applicable to atomic and molecular complexes, 
as developed by Berrington and co-workers  \citep{qb1,qb2,qb3} 
was used to locate and determine the resonance positions. 
The resonance width $\Gamma$ may also
be determined from the inverse of the energy derivative
of the eigenphase sum $\delta$ at the resonance energy $E_r$ via
\begin{equation}
	\Gamma = 2\left[ \frac{d\delta}{dE}\right]^{-1}_{E=E_r} = 2\left[\delta^{\prime}\right]^{-1}_{E=E_r} .
\end{equation}
	%
	%
	%
\begin{figure*}
\includegraphics[width=13.5cm]{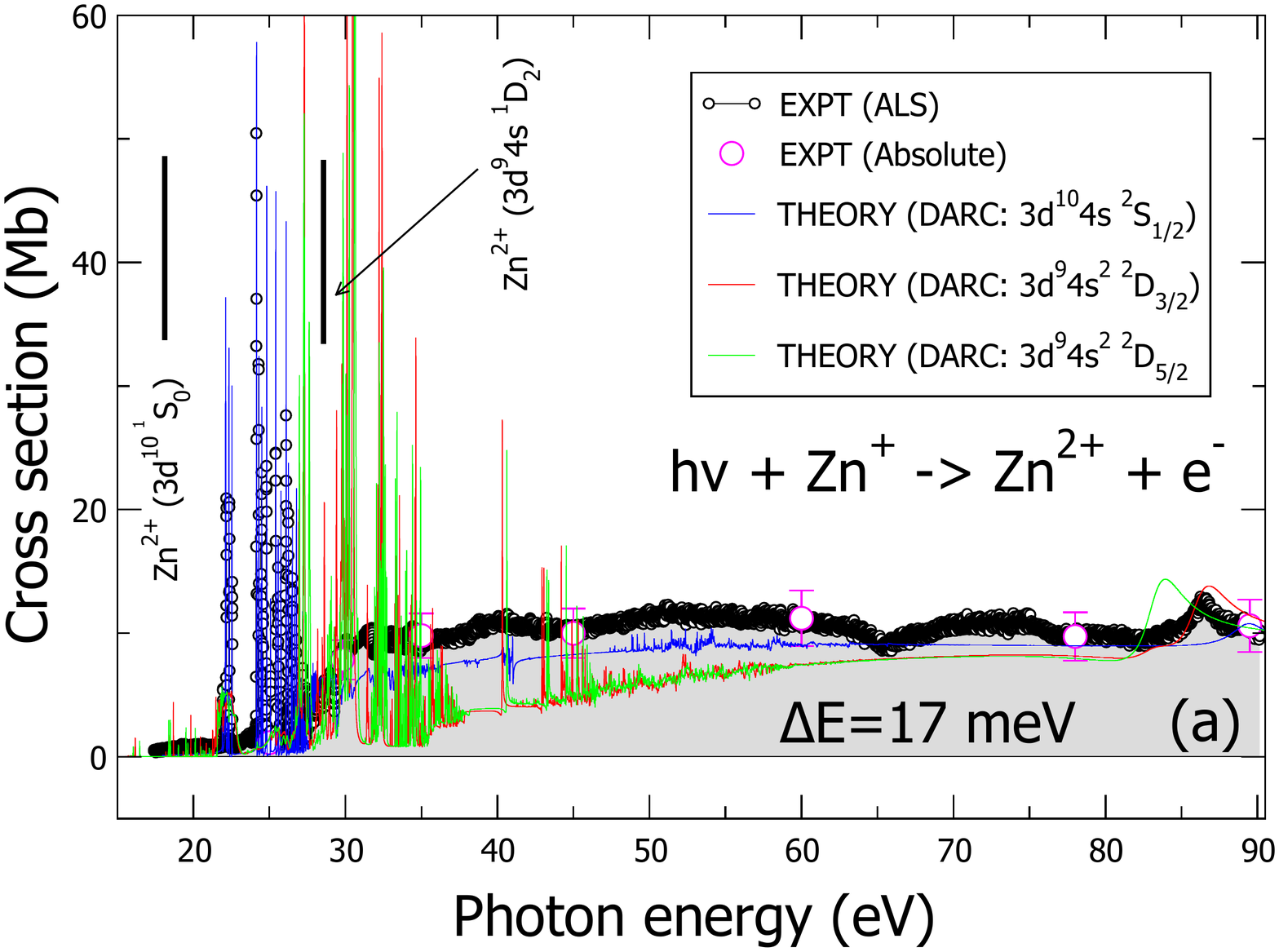}
\vspace*{0.1cm}
\includegraphics[width=13.5cm]{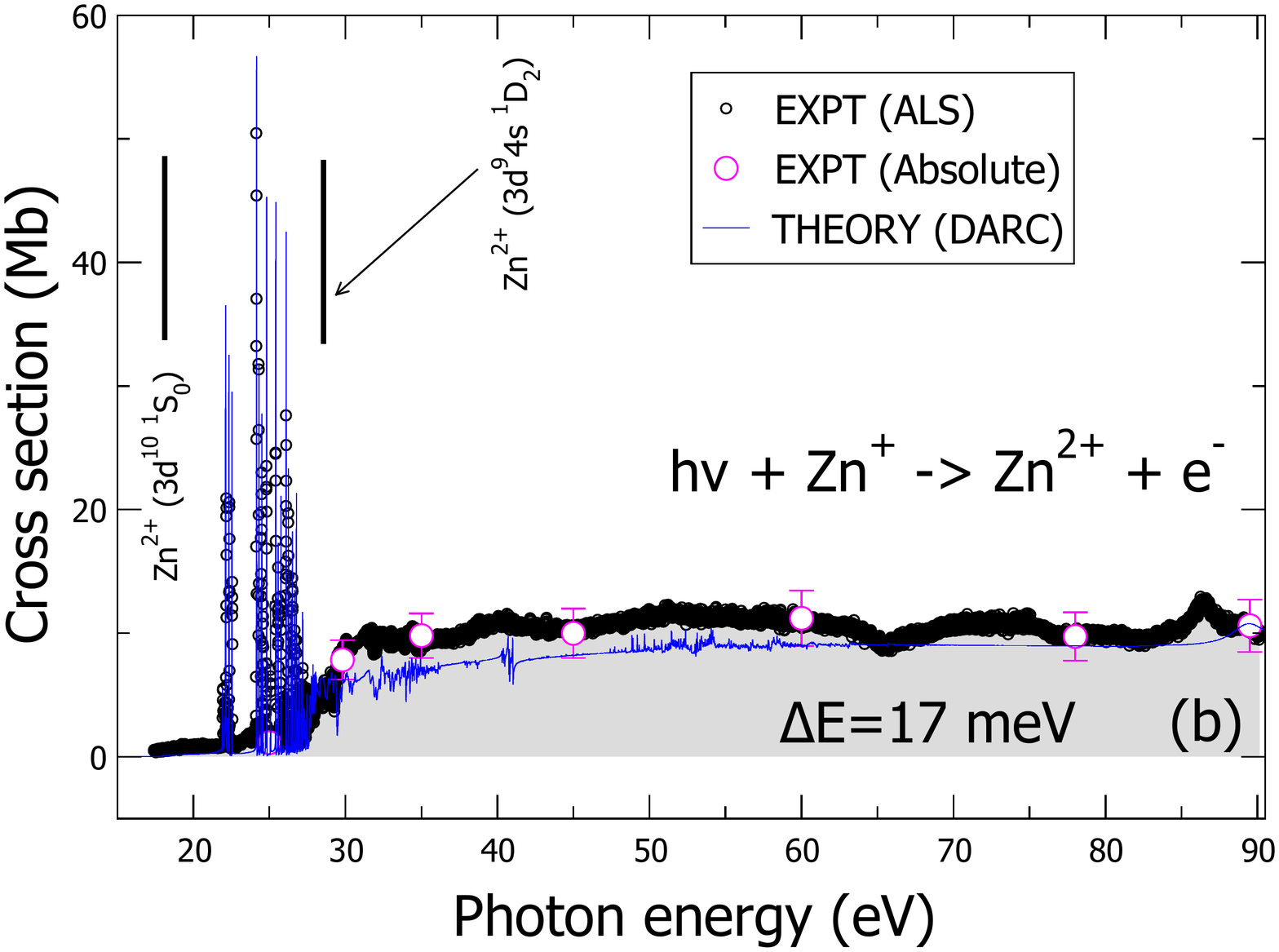}
\caption{(Colour online) Single photoionization cross section of Zn$^{+}$ : 
         ALS experimental data (open circles) and the 
         355-level Dirac $R$-matrix theoretical results.  
         The ALS spectrum consists of Zn$^{2+}$ ion yield spectra 
         normalized to the absolute cross section measurements 
         (magenta open circles).   The thresholds for 
          Zn$^{2+}(3d^{10}\;{\rm ^1S}_0)$ at 17.964 eV  and 
         Zn$^{2+}(3d^94s\; {\rm ^1D}_2)$ at 28.317 eV, 
         respectively are indicated by vertical lines. 
         The correction factor $f_c$ has not been applied to the 
         ALS data illustrated in the figures.  (a) ALS measurements and the
         DARC PI calculations for the ground ($3d^{10}4s\;{\rm ^2S}_{1/2}$) (blue line) and 
         metastable ($3d^94s^2\;{\rm ^2D}_{3/2}$ (red line) and $3d^94s^2\;{\rm ^2D}_{5/2}$) 
         (green line)  contributions.
         (b) ALS measurements and the DARC PI calculations with an appropriate mixture of 
         the ground and metastable states. See text for details}
\label{fig1}
\end{figure*}

	%
	%
\begin{figure*}
\includegraphics[width=13.5cm]{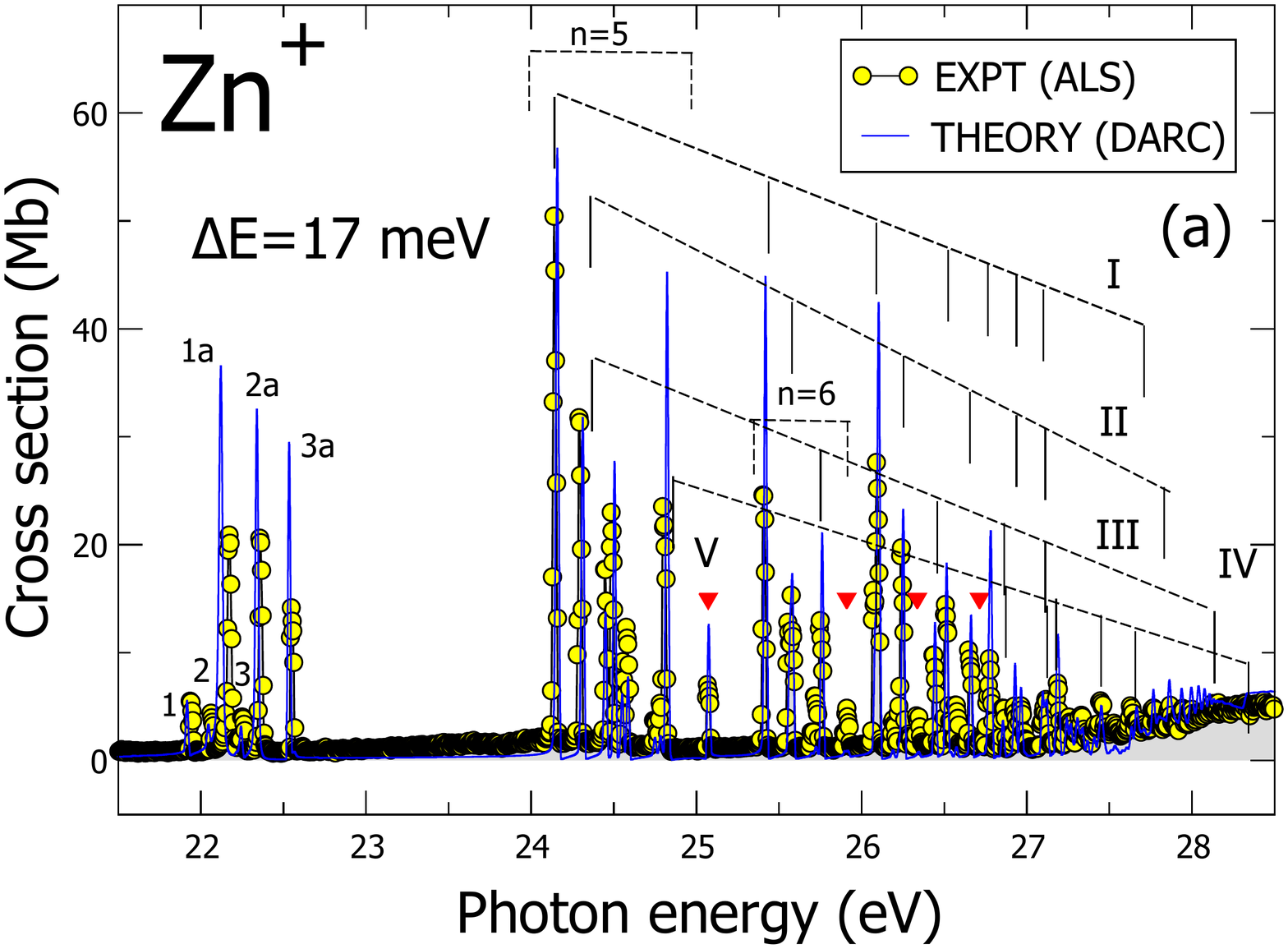}
\vspace*{0.1cm}
\includegraphics[width=13.5cm]{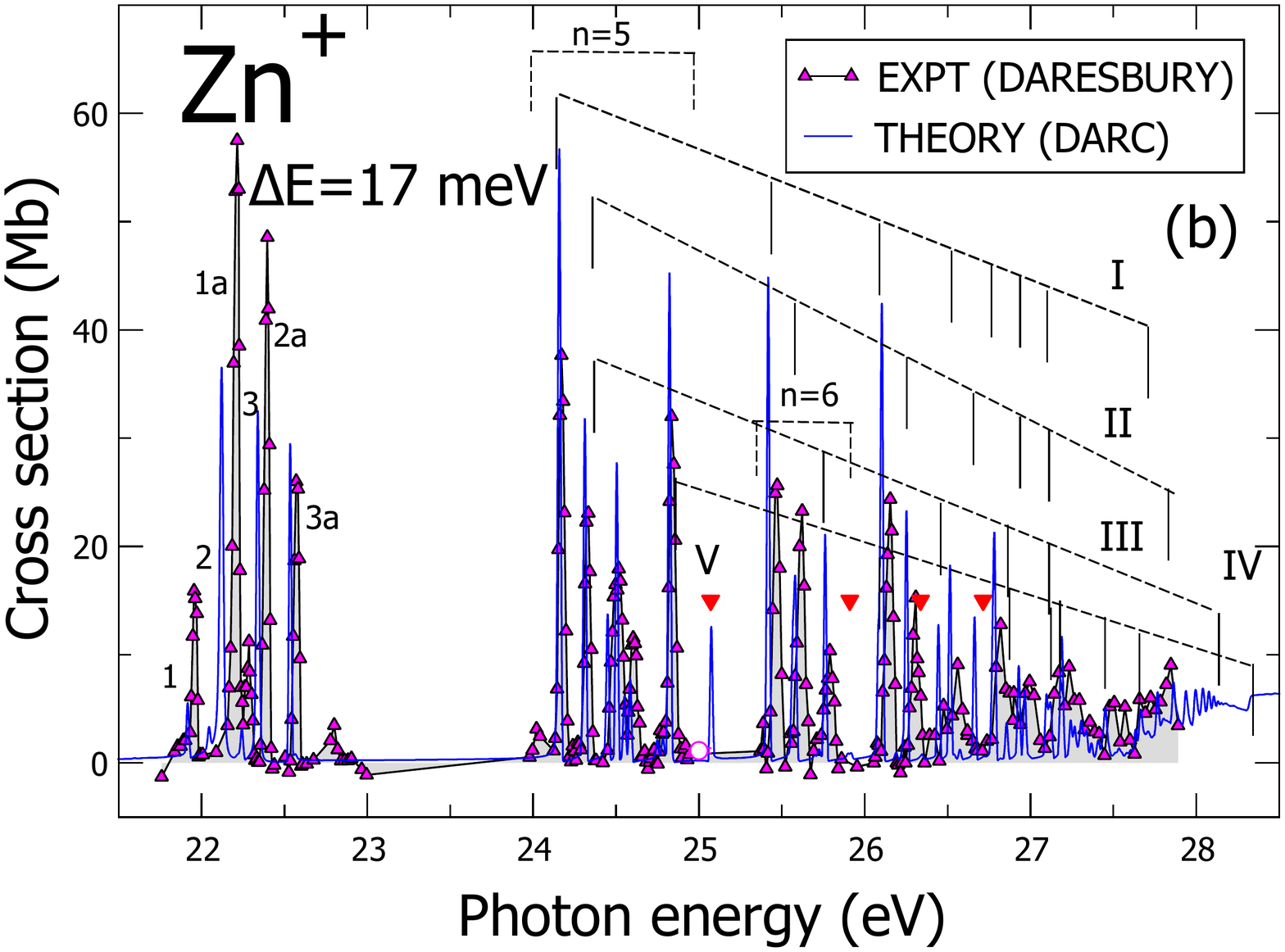}
\caption{(Colour online) Single-photon, single photoionization cross section of 
         Zn$^{+}$: Resonances features in the spectrum for 21.5 -- 28.5 eV.  
         (a) ALS measurements solid circles (yellow), resolution of 17 meV FWHM. 
         (b) DARESBURY solid triangles (yellow) \citep{Peart1987}.  
         In both panels the large-scale DARC calculations solid line (blue) are convoluted 
         with a Gaussian, having a profile of 17 meV FWHM, with an appropriate admixture of 
         ground and metastable states, see text for details.
         Vertical lines indicate the resonant energies of Table \ref{tab1}. The final vertical 
         lines to the right of each group correspond to the limit of the series. 
         Gaussian fits to the peak centres are tabulated in Tables \ref{tab1} and \ref{tab2}.
         The grouped lines of the Rydberg series and vertical lines from panel (a) for 
         those resonances that can be paired with a peak in the data by \citet{Peart1987}. 
         Open circle (magenta) is the ALS  absolute cross section at 25 eV.} 
\label{fig2}
\end{figure*}
	%
	%
\begin{figure*}
\includegraphics[width=13.5cm]{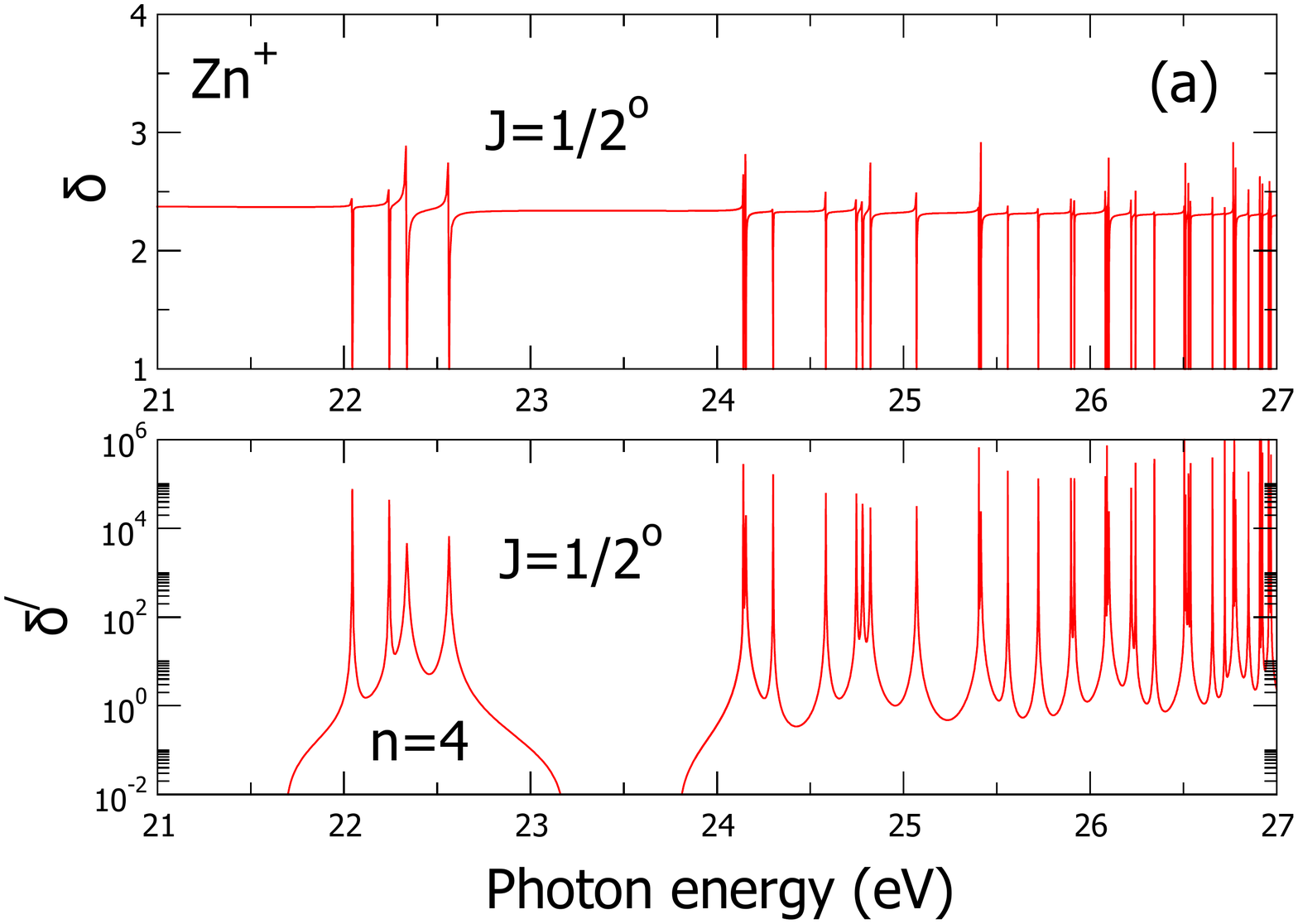}
\vspace*{0.1cm}
\includegraphics[width=13.5cm]{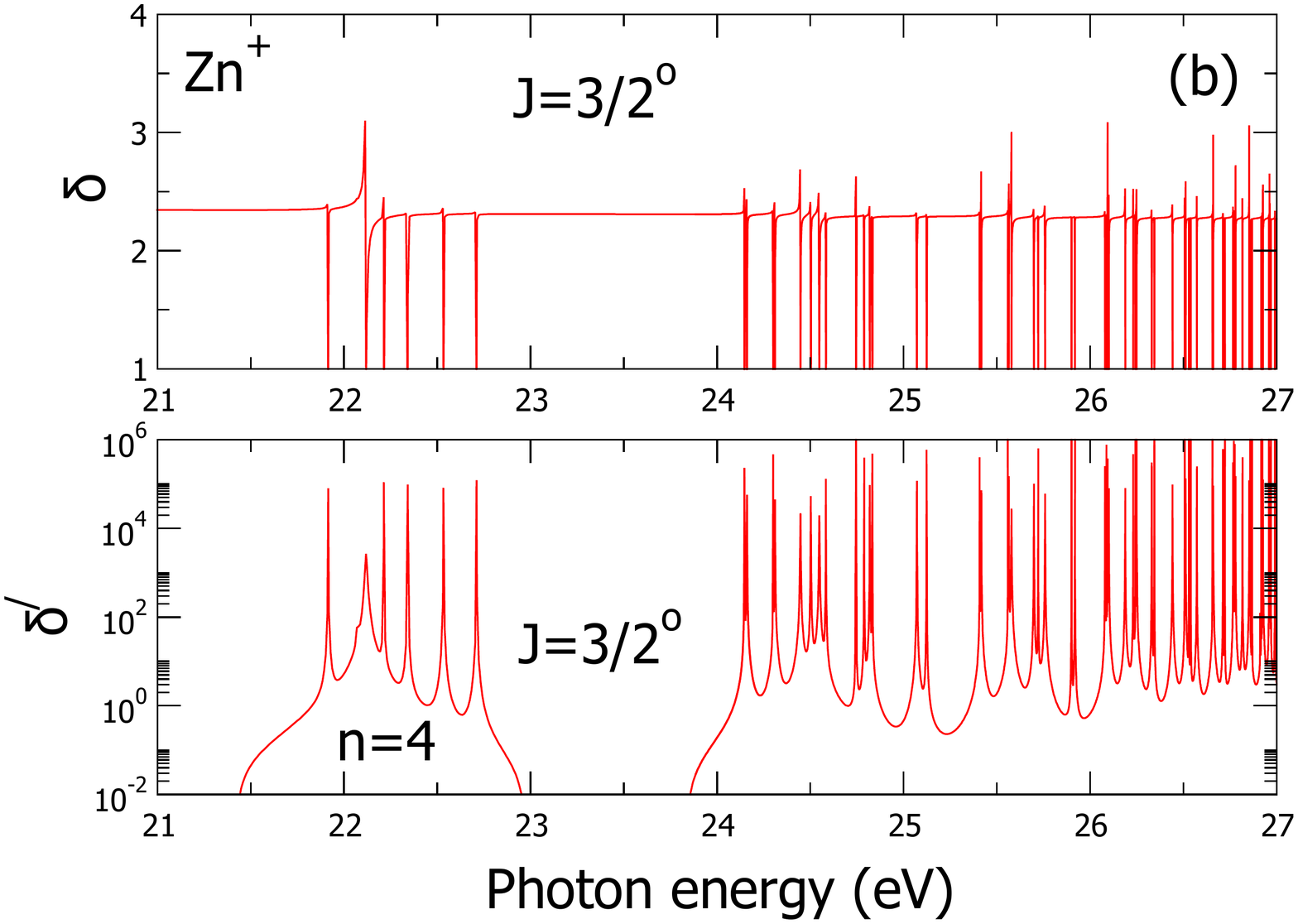}
\caption{(Colour online) Eigenphase sum $\delta$ and its derivative $\delta^{\prime}$ in the photon energy
                 region 21 - 27 eV for (a) J=1/2 odd scattering symmetry, (b) J=3/2 odd scattering symmetry
                 from the large - scale DARC PI calculations on the Zn II ion. 
                Interloping and overlapping resonances series 
                in the Zn II spectrum disrupt the 
                regular Rydberg resonance series pattern. Tables \ref{tab3} and \ref{tab4} 
                give assignments of resonances for each scattering symmetry.}
\label{fig3}
\end{figure*}
\begin{figure*}
\includegraphics[scale=1.50,width=18.0cm]{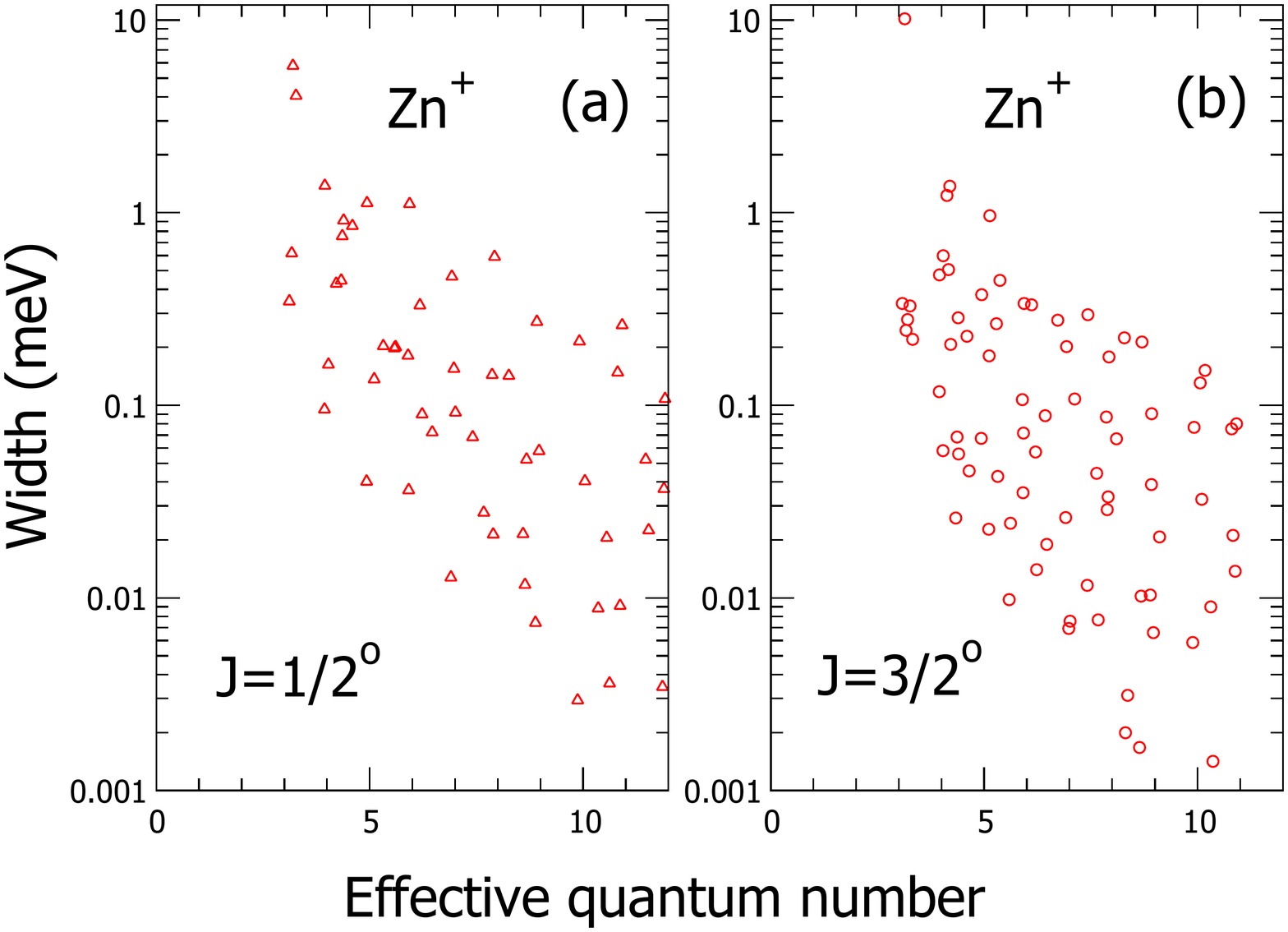}
\caption{(Colour online) Resonance line widths $\Gamma$ (meV) versus the effective quantum number $\nu$ 
                for resonances series found in the Zn II ALS spectrum in the photon energy 
                range 21 --  27 eV.  (a) J=1/2 odd symmetry, (b) J=3/2 odd symmetry. 
                The theoretical values were obtained using the eigenphase derivative method
                \citep{qb1,qb2,qb3} from the large-scale DARC calculations on the Zn II ion. 
                The resonanances line widths are 10 meV or much less and impossible 
                to be extracted with the present limited photon resolution of 17 meV FWHM.}
\label{fig4}
\end{figure*}

\section{Results and Disscussion}

The ALS spectrum for the single photoionization of Zn$^{+}$ cross section measurements 
are illustrated in Fig. \ref{fig1} (a) and (b), along with the absolute measurements and  the results 
from the large-scale DARC PI cross section calculations.
The ALS measurements were made in the photon energy range 17.5 eV -- 90.0 eV with a 
photon energy resolution of 17 meV. In order to compare directly with 
experiment, the DARC results have been convoluted 
with a Gaussian having a profile width of 17 meV. PI calculations 
were performed for the ground state $3d^{10}4s\;{\rm ^2S}_{1/2}$ 
and the  $3d^94s^2\;{\rm ^2D}_{3/2,5/2}$ metastable states. 
 The absolute PI cross section measurements are shown 
 in Fig. \ref{fig1} (a) and (b) with open circles.   
Note, the correction factor $f_c$ has not been applied to the ALS measurements. 
The DARC PI calculations are included in Fig. \ref{fig1} (a) for 100\% population of both 
the ground and metastable states.  
Fig. \ref{fig1} (b) shows a comparison of the DARC PI cross section, 
convoluted using a Gaussian having a profile of 17 meV and an appropriate 
weighted admixture of 98\% ground and 2\% of the 
statistical average of the metastable states, with the ALS measurements. 
Excellent agreement is obtained between theory and experiment over the energy 
region investigated.

A restricted region of the spectrum, for the photon energy range  21.5 -- 28.5 eV,  
 is displayed in Fig. \ref{fig2} (a) and (b).  Fig. \ref{fig2} (a) shows the
present ALS spectrum recorded at 17 meV FWHW compared to 
the DARC calculations.  In Fig. \ref{fig2}  (b)   
the previous Zn II digitized spectra obtained at the Daresbury 
radiation facility \citep{Peart1987}  are illustrated 
for the same energy region with a similar comparison made 
with the DARC calculations.   
In Figs. \ref{fig2} (a) and (b), 
numerous sharp peak structures are seen in the Zn II spectrum over
this energy region.  The peaks in the spectrum can be  
assigned to well-defined Rydberg series originating from the ground state, 
namely, Zn$^{+}(3d^{10}4s \; {\rm ^2S}_{1/2})$ resonances, 
converging to the triplet state $3d^94s\; {^3D}_{3,2,1}$  thresholds 
(i.e. series I, II and III in panel (a) of Fig. \ref{fig2} ), and to the 
singlet state threshold $3d^94s\; {\rm ^1D_2}$ (series IV) of the Zn$^{2+}$ ion.

Series identification was guided by fitting 
equation (\ref{rydberg}) to a particular set of peak 
positions that corresponded to a well-defined series.
The Rydberg series  limits  $E_{\infty}$ were determined 
by fitting the series with equation (\ref{rydberg}),  
which agreed with results from the tabulations in the 
NIST database \citep{NIST2016}, to three significant figures.   
Due to several overlapping peaks, values for the resonance energies 
were derived from Gaussian fits to the peaks to obtain more precise values for the peak centroids. 
These centroids are tabulated in Table \ref{tab1}, together with the averaged quantum 
defects $\mu_f$ and the corresponding series limits for the identified Rydberg resonance series. 

In the resonant region of the spectrum 24 - 28 eV, it is  seen that there are 
many interloping and overlapping resonances present. For some of the peaks, their  
resonance energies in Table \ref{tab1} are given in parenthesis (or brackets) 
in order to indicate that they are estimates. 

A resonance series with $n \ge $ 6, labeled as series V in panel (a) of Fig. \ref{fig2},
(having small strengths), with the  same series energy 
limit  $E_{\infty}$=27.647 eV of series I can be identified in the experimental spectrum. 
Excitation energies were calculated using the 
Cowan code \citep{Cowan1981,Mann1983,Clark1991}. 
To compare them with the measured resonance energies $E_n$ (see Fig. \ref{fig3}),  
a linear fit was used where
\begin{equation}
 	E_n= m \; E_{\rm Cowan} + b
\label{cow}
\end{equation}
 with $m$ and $b$ constants \citep{Mueller2014}. This function fitted the measured 
resonance energies very well for the excited state $E_{\rm Cowan}$  resonance energies. 

Below 23 eV, a group of well-resolved resonances is observed, as indicated in panel (a) of Fig. \ref{fig2}. 
Higher-order radiation in the photon beam appears in the low energy interval of the 
first grating of the photon beamline, which may result in excitation of the Zn$^+$ 
energy levels accessible at higher photon energies. 
If this is the case, one would expect resonances to also 
appear at two or three times their photon energies. 
Since no such peaks were observed (see Fig.\ref{fig1} (a) and (b)) 
at the corresponding higher photon energies, 
one may conclude that higher-order photons are not the 
source of this particular set of resonances.  

%
%
%
	%
 \begin{table*}
  \caption{Resonance energies $E_n$ of the identified 
		Rydberg series I, II, III, IV and V (from the present experimental ALS data) 
 		converging to the $3d^{9}4s\;{\rm ^3D}_{3,2,1}$ triplet and the 
 		$3d^{ 9}4s\;{\rm ^1D}_2$ singlet states originating from the Zn$^+$ ground state. Some 
 		energies, in brackets, have the same superscript to indicate that they 
 		superimpose. In some cases, multi-peak Gaussian fits 
	 	were used to try to improve assignments when possible. Roman number labels 
 		correspond to those of Fig. \ref{fig2}. 
        The average value of the quantum defect $\mu_f$ for the Rydberg series 
		corresponds to the value of $\mu$ derived from direct fitting to equation (\ref{rydberg}).
 		Values in parenthesis or brackets are estimates from the fits. \label{tab1}}
\begin{center} 
 \begin{tabular}{ccccccccccc}
 \hline \hline \\  [-2.0ex]
        	& &   \multicolumn{9}{c}{Autoionizing Zn II Rydberg resonance series energies}           \\ 
[0.5ex]
                                           \cline{3-11}                      \\ 
[-1.75ex]
        	& &           I             				& &           II            	
	& &           III           				& &           IV            
	& &           V            \\ 
[0.5ex]                                           
        	& & $(3d^{ 9}4s){\rm ^3D}_3 \ np$ 	& & $(3d^{ 9}4s){\rm ^3D}_2 \ np$ 
	& & $(3d^{ 9}4s){\rm ^3D}_1 \ np$ 	& & $(3d^{ 9}4s){\rm ^1D}_2 \ np$ 		
	& & $(3d^{ 9}4s){\rm^3D}_3 \ np$\\ 
[0.5ex]
       \cline{3-3}                    \cline{5-5}       \cline{7-7}      \cline{9-9}  \cline{11-11}       \\ 
[-3ex]
   \\ 
[-1.75ex] 
  $n$   & & $E_n$(eV)  	& & $E_n$(eV)    	&  &$E_n$(eV)  	& &$E_n$(eV)   	& & $E_n$(eV)\\ 
\cline{1-1}       \cline{3-3}                    \cline{5-5}       \cline{7-7}      \cline{9-9}  \cline{11-11}              \\                  
   5    & &  24.141    	& &   24.294    	&  &  24.488   	& & 24.804    	& &   --            \\
   6    & &  25.408    	& &   25.569    	&  &  25.752   	& &[26.069]$^1$	&  & 25.071     \\
   7    & &[26.093]$^1$	& &   26.243    	&  &  26.439   	& &[26.773]$^2$	&  & 25.910     \\
   8    & &  26.512    	&  &  26.656    	&  &  26.858  	& &[27.184]$^4$	&  &(26.340)    \\
   9    & &[26.773]$^2$     &  &  26.926    	&  &[27.116]$^3$	& &  27.450    	&  &(26.714)    \\
   10    & &  26.966            &  &[27.104]$^3$ 	&  &    --     		& &  27.632    	&  &   --            \\
   11   & &[27.087]$^3$    &  &     --      		&  &    --     		& &  27.767    	&  &   --            \\
   12   & &[27.184]$^4$    &  &     --      		&  &    --     		& &  27.863    	&  &   --            \\
   13   & & (27.253)            &  &     --      		&  &    --     		& &  27.939    	&  &   --            \\
   14   & &     --                   &  &     --      		&  &    --     		& &  27.994    	&  &   --            \\
   15   & &     --                   	&  &     --      		&  &    --     		& &  28.041    	&  &   --            \\
 \vdots & &            		&  &             		&  &           		& &            		&  &                  \\
  Limit & &  27.647    	&  &   27.793    	&  &  27.989   	& &  28.317    	&  & 27.647     \\ 
$\mu_f$&& 1.11 $\pm$ 0.2 &  &    1.08 $\pm$ 0.2  &  &1.07 $\pm$ 0.2  & & 1.05 $\pm$ 0.2    &  &  1.43 $\pm$ 0.3 \\     
 \hline \hline                                                               
 \end{tabular} 
 \end{center}
 \end{table*}
\begin{figure*}
\includegraphics[scale=1.50,width=18.0cm]{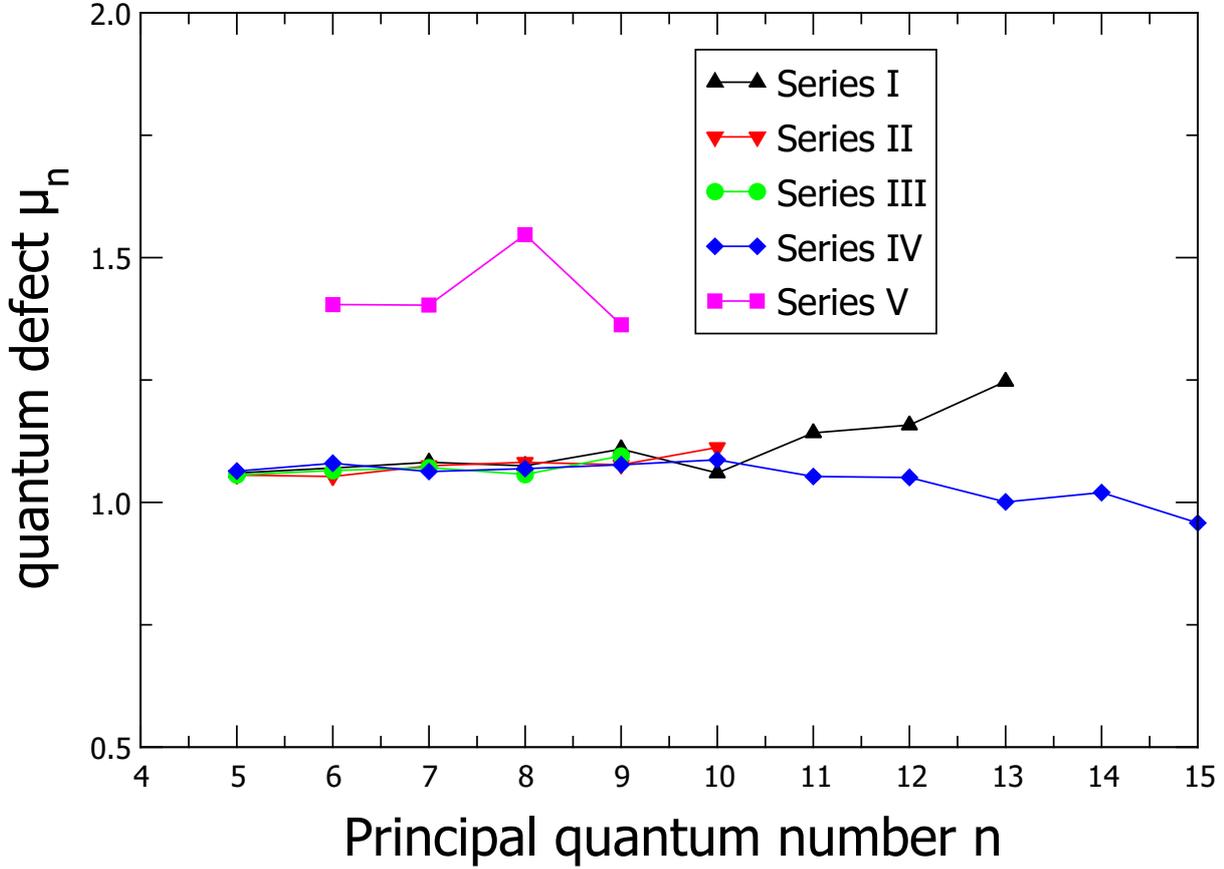}
\caption{(Colour online) Quantum defect $\mu_n$ versus the principal quantum number $n$ 
        for the dominant resonances series found in the Zn II ALS measured spectrum for the photon energy 
        range 24 --  28 eV. Table \ref{tab1} gives details for resonance series labeled I, II, III, IV and V. }
\label{fig5}
\end{figure*}
 	%
%
%
%
%
 \begin{table*} 
 \caption{Resonance energies ($E_n$) and strengths ($\overline{\sigma}_{\rm PI}$) from the ALS data  
		 only for those peaks  resolved in the DARESBURY data of \citet{Peart1987} where
 		a comparison is possible. The resonance strengths $\overline{\sigma}_{\rm PI}$ derived 
 		from the present ALS data have been multiplied by $f_c$. 
        Resonance strengths and positions were determined from the large-scale  355-level
		DARC calculations, assuming 100\% of the ground state. 
        The resonance positions for the DARC values were 
        estimated by fitting the peaks in the total cross sections to Fano profiles. 
        Values in brackets shown (98\% of the ground state)  are corrected
        for metastable contamination  in the ion beam, 
        determined to be 2\%, see text for details.\label{tab2}}
 \begin{center}
 \begin{tabular}{ccccccccccccc} 
\hline\hline \\[-1.0ex]
 Resonance	& & \multicolumn{3}{c}{ ALS$^a$  } 
		& & \multicolumn{3}{c}{DARESBURY$^b$} 
		& & \multicolumn{3}{c}{DARC$^c$}\\
\cline{1-1}    	\cline{3-5}                  \cline{7-9}              \cline{11-13} \\
        		& & $E_n$ 		& &    $\overline{\sigma}_{\rm PI}$    
        		& & $E_n$ 		& &    $\overline{\sigma}_{\rm PI}$     		
                     & & $E_n$     	& &    $\overline{\sigma}_{\rm PI}$\\    
 Label           	& &   (eV)  		& & ($Mb \ eV$)  
	 	& &   (eV)  		& &  ($Mb \ eV$)	
		& &   (eV)  		& &  ($Mb \ eV$)\\
\cline{1-1}        \cline{3-5}                 \cline{7-9}  \cline{11-13} \\
$n = 4$	&&			&&				&&		&&		&&		&&		   \\
  1   (J=3/2)  & &  21.940 		&  &    0.19   $\pm$  0.05 	& & 21.958 	& &  0.55   	&& 21.919	&& 0.19 (0.18) \\
  1a (J=3/2) 	& &  22.173		&  &    0.87   $\pm$ 0.21 	& & 22.210 	& &  2.50   	&& 22.123	&& 1.22 (1.20) \\
  2a (J=1/2)	& &  22.361 		&  &    0.69   $\pm$ 0.20 	& & 22.393 	& &  1.70   	&& 22.340	&& 0.72 (0.71) \\
  3a (J=3/2)  & &  22.552 		&  &    0.42   $\pm$ 0.13 	& & 22.573 	& &  0.98    	&& 22.535	&& 0.45 (0.44) \\  
\cline{1-1}         \cline{3-5}                  \cline{7-9}  	\cline{11-13}   \\
$n = 5$ 	& &  24.141 		&  &    1.52   $\pm$ 0.50 	& & 24.172 	& &  1.60  	&& 24.158	&& 1.42 (1.39) \\
        		& &  24.294 		&  &    0.99   $\pm$ 0.31	& & 24.329 	& &  0.98 	 && 24.312	&& 0.83 (0.81) \\
        		& &  24.488 		&  &    0.64   $\pm$ 0.23 	& & 24.509 	& &  1.36  	&& 24.504	&& 0.53 (0.52) \\
        		& &  24.804 		&  &    0.84   $\pm$ 0.24 	& & 24.839 	& &  1.57  	&& 25.824	&& 0.97 (0.95) \\
\cline{1-1}       \cline{3-5}                  \cline{7-9}  \cline{11-13} \\
$n = 6$ 	& &  25.408 		& &    0.77   $\pm$ 0.25  	& & 25.468 	& &  1.34 	&& 25.418	&& 1.04 (1.02) \\         
        		& &  25.569 		& &    0.75   $\pm$ 0.15  	& & 25.616 	& &  1.24 	&& 25.577	&& 0.64 (0.63) \\         
        		& &  25.752 		& &    0.35   $\pm$ 0.13  	& & 25.786 	& &  0.70 	&& 25.762	&& 0.41 (0.40) \\ 
\cline{1-1}      \cline{3-5}                  \cline{7-9}   \cline{11-13} \\
$n = 7$ 	& & (26.093)		&  &    (1.13  $\pm$ 0.28)  & & 26.150 	& &  1.28 	&&26.102 	&& 1.32 (1.29) \\        
        		& & 26.243  		&  &     0.62  $\pm$  0.20  	& & 26.305 	& &  0.77  	&&26.251 	&& 0.65 (0.84) \\        
 \hline\hline \\
 \end{tabular}
 \end{center}
\begin{flushleft}
$^a$ALS, present experiment.\\
$^b$DARESBURY, previous experimental results \citep{Peart1987}.\\
$^c$DARC, present theory, 355-level approximation.\\
\end{flushleft}
\end{table*}
 \begin{table*}
  \caption{Resonance energies $E_n$ of the identified Rydberg 
           series I, II, III, IV and V (from the large-scale DARC calculations), 
           J=1/2 odd scattering symmetry, 
 		   converging to the $3d^{9}4s\;{\rm ^3D}_{3,2,1}$ triplet and the 
 		   $3d^{ 9}4s\;{\rm ^1D}_2$ singlet states originating from 
           the Zn$^+$ ground state.\label{tab3}}
\begin{center} 
 \begin{tabular}{ccccccccccc}
 \hline \hline \\  [-2.0ex]
        	& &   \multicolumn{9}{c}{Autoionizing Zn II Rydberg resonance series energies, J=1/2$^{o}$ symmetry}           \\ 
[0.5ex]
                                           \cline{3-11}                      \\ 
[-1.75ex]
        	& &           I             				& &           II            	
	& &           III           				& &           IV            
	& &           V            \\ 
[0.5ex]                                           
        	& & $(3d^{ 9}4s){\rm ^3D}_3 \ np$ 	& & $(3d^{ 9}4s){\rm ^3D}_2 \ np$ 
	& & $(3d^{ 9}4s){\rm ^3D}_1 \ np$ 	& & $(3d^{ 9}4s){\rm ^1D}_2 \ np$ 		
	& & $(3d^{ 9}4s){\rm^3D}_3 \ np$\\ 
[0.5ex]
       \cline{3-3}                    \cline{5-5}       \cline{7-7}      \cline{9-9}  \cline{11-11}       \\ 
[-3ex]
   \\ 
[-1.75ex] 
  $n$   & & $E_n$(eV)  	& & $E_n$(eV)    	&  &$E_n$(eV)  	& &$E_n$(eV)   	& & $E_n$(eV)\\ 
\cline{1-1}       \cline{3-3}                    \cline{5-5}       \cline{7-7}      \cline{9-9}  \cline{11-11}              \\
   4    &&	--		&  & 22.044	&  &	--		& & 22.337	& & 22.563	\\                  
   5    & & 24.149     	&  & 24.583  &  &  24.141   	& & 24.747   	& & 24.780    \\
   6    & & 25.415  		&  & 25.559  &  &  25.404   	& & 25.723	& & 25.723     \\
   7    & & 26.101		&  & 26.220  &  &  26.082   	& & 26.345	& & 26.345    \\
   8    & & 26.512		&  & 26.538  &  &  26.504   	& & 26.656	& & 26.659    \\
   9    & & 26.780	    	&  & 26.849  &  &  26.773		& & 26.909   	& & 26.910    \\
 10    & & 26.962               	&  & 26.147  &  &  26.957     	& & 27.088   	& & 27.088    \\
  \vdots & &            		&  &             	&  &                	& &               & &                 \\
  Limit & &  27.647    	&  & 27.793  &  &  27.989   	& &  28.317 &  & 27.647     \\ 
 \hline \hline                                                               
 \end{tabular} 
 \end{center}
 \end{table*}
 \begin{table*}
  \caption{Resonance energies $E_n$ of the identified 
		   Rydberg series I, II, III, IV and V (from the large-scale DARC calculations), 
           J=3/2 odd scattering symmetry, 
 		   converging to the $3d^{9}4s\;{\rm ^3D}_{3,2,1}$ triplet and the 
 		   $3d^{ 9}4s\;{\rm ^1D}_2$ singlet states originating from the Zn$^+$ ground state.
           \label{tab4}}
\begin{center} 
 \begin{tabular}{ccccccccccc}
 \hline \hline \\  [-2.0ex]
        	& &   \multicolumn{9}{c}{Autoionizing Zn II Rydberg resonance series energies, J=3/2$^{o}$ symmetry}           \\ 
[0.5ex]
                                           \cline{3-11}                      \\ 
[-1.75ex]
        	& &           I             				& &           II            	
	& &           III           				& &           IV            
	& &           V            \\ 
[0.5ex]                                           
        	& & $(3d^{ 9}4s){\rm ^3D}_3 \ np$ 	& & $(3d^{ 9}4s){\rm ^3D}_2 \ np$ 
	& & $(3d^{ 9}4s){\rm ^3D}_1 \ np$ 	& & $(3d^{ 9}4s){\rm ^1D}_2 \ np$ 		
	& & $(3d^{ 9}4s){\rm^3D}_3 \ np$\\ 
[0.5ex]
       \cline{3-3}                    \cline{5-5}       \cline{7-7}      \cline{9-9}  \cline{11-11}       \\ 
[-3ex]
   \\ 
[-1.75ex] 
  $n$   & & $E_n$(eV)  	& & $E_n$(eV)    	&  &$E_n$(eV)  	& &$E_n$(eV)   	& & $E_n$(eV)\\ 
\cline{1-1}       \cline{3-3}                    \cline{5-5}       \cline{7-7}      \cline{9-9}  \cline{11-11}             \\
   4    &&	--		&&  22.214		&  &  21.916		& & 22.213		&  & 22.711   \\                  
   5    & & 24.148     	& & 24.311    	&  &  24.447   	& & 24.503    	&  & 24.745   \\
   6    & & 25.417  		& & 25.573    	&  &  24.564     	& & 25.579		&  & 25.759   \\
   7    & & 26.101		& & 26.243    	&  &  26.233   	& & 26.330		&  & 26.330   \\
   8    & & 26.512		&  &26.656    	&  &  26.572  	& & 26.656		&  & 26.572   \\
   9    & & 26.780	    	&  &26.858    	&  &  26.853		& & 26.868 		&  & 26.817   \\
 10    & & 26.963               	&  &26.991 		&  &    --     		& & 26.991 		&  & 26.991   \\

 \vdots & &            		&  &             		&  &           		& &            		&  &               \\
  Limit & &  27.647    	&  &   27.793    	&  &  27.989   	& &  28.317    	&  & 27.647   \\ 
 \hline \hline                                                               
 \end{tabular} 
 \end{center}
 \end{table*}

In panel (b) of Fig. \ref{fig2} the earlier findings of \citet{Peart1987} 
are illustrated along with the DARC results.   
In this figure, as a guide, we have retained some of the grouping lines, labels, 
and a single absolute cross section data point from panel (a) of Fig. \ref{fig2}. 
The peaks reported by \citet{Peart1987} appear shifted toward higher energies, 
or shorter wavelengths, compared to the present ALS measurements and the DARC calculations. 
Differences are seen ranging from about 20 meV (in the lower energy regime), 
to 60 meV (in the higher energy regime)  with their experiment. These 
differences we attribute to the energy calibration 
uncertainties reported by \citet{Lyon1986} of 30 meV to 77 meV,
 in the respective energy regimes. We note also that the DARC peak energies 
favour the ALS measurements as can be seen from Fig. \ref{fig2} (a), (b) and 
the results tabulated in Table \ref{tab2}.

The peaks in the Zn II spectrum were identified from 
a theoretical analysis of the eigenphase sum $\delta$ and its derivative $\delta^{\prime}$ 
for the J=1/2 and J=3/2 odd scattering symmetries obtained from the 
large-scale DARC calculations.  The results are presented in Figs. \ref{fig3} (a) and (b).
 We find that the peaks below 23 eV originate from photoionization of the ground state term
and attribute them to excited resonance  states 
$[3d^{ 9}4s\;({\rm ^{1,3}D}_{3,2,1})\ 4p]_{1/2,3/2}$ converging to different 
Zn$^{2+}$ core states. These are labeled, 1a -- 3a and 1 -- 3 in the experimental spectrum. 
The remaining peaks we can assign to well-defined Rydberg resonance series.
We note that the $3d^{ 9}4s\;({\rm ^3D}_3)\ 5p$ resonance, identified 
with series I, located at 24.141 eV, is the strongest in the spectrum. 

An important feature of the spectrum of Fig.\ref{fig1} (a) and (b) 
is the small non-zero cross section below 
the ionization threshold (marked by vertical line). 
This shift is probably due to the presence of higher-order radiation 
in the photon beam which causes a contribution 
from the non-resonant cross section or from metastable contamination of the ion beam. 

To determine the metastable contaminant in the ion beam, large-scale  
DARC  PI calculations were performed for the 
$3d^94s^2\;{\rm ^2D}_{3/2,5/2}$ initial metastable states.
From these DARC PI cross section calculations we determined that the metastable 
contamination of the ion beam was 2\% by using a statistical average 
of the metastable theoretical cross sections 
compared to the ALS measurements at 17.5 eV.  
Hence we conclude that the resonance features  
in the spectrum originate from the ground state.  Fig. \ref{fig1} (a) 
shows the ALS measurements and the DARC PI calculations for the 
ground and metastable states, where 100\% population of the 
ground and metastable states are assumed.     
In Fig. \ref{fig1} (b), the DARC PI calculations have 
been appropriately weighted, 98\% ground state and 2\% of the statistical
average of the metastable states. As can be 
seen from the results presented in Fig. \ref{fig1} (a) and (b) 
there is excellent agreement between experiment and theory. 

We note that in some  cases, the populations of  
excited states in the ion beams have been estimated to be relatively low \citep{Mueller2014}.
For example, in a similar ion source, for C-like ions, using the MCDF method, 
the population of the $\rm ^5S_2$ level was estimated 
to be 2 -- 3\% \citep{Bizau2005}.   In the present work  by 
performing PI calculations for the metastable states indicated above 
we find their contribution is small, around 2\%.

The prominent Rydberg resonance series are identified as Zn$^+(3d^94s\; np)$ excitations,  
as indicated by the series labeled I, II, III and IV in panel (a) of Fig. \ref{fig2}..  
Resonance energies $E_n$ of each Rydberg resonance 
series found in the ALS spectrum are tabulated in Table \ref{tab1} 
along with the averaged quantum defect $\mu_f$ for each series. 
A Zn$^+(3d^94s\; np)$, $n \ge $ 6,  Rydberg resonance series, as  
indicated by the downward pointing solid triangles (red), 
see Figs. 2 (a) and (b), was also identified in the spectrum.
The widths of many of these resonances found in the spectrum, 
as illustrated in Fig. \ref{fig4} (a) and (b), 
are 10 meV or less  making it impossible to extract values 
with the present limited experimental resolution of 17 meV.
 The quantum defects $\mu_n$ as a function of the 
principal quantum number $n$, for the dominant series are illustrated for the 
five series, I, II $\dots$, V, in Fig. \ref{fig5}.

An additional check on the theoretical data was carried out by comparing  
the integrated continuum oscillator strength $f$ with experiment. 
The integrated continuum oscillator strength $f$ of the experimental spectra was 
calculated over the energy grid [$E_1$, $E_2$], where $E_1$ is the minimum experimental energy  
and $E_2$ is the maximum experimental energy measured, (respectively 17.409 eV and 91.142 eV),  
using  \citep{Shore1967,Fano1968,berko1979},
\begin{eqnarray*}
f  &  = & 9.1075 \times 10^{-3} \int_{E_1}^{E_2} \sigma (h\nu) dh\nu  \nonumber \\ 
   & =  & 9.1075 \times 10^{-3} \; \; \overline{\sigma}_{\rm PI}   \\
\end{eqnarray*}
where 
\begin{equation}
 \overline{\sigma}_{\rm PI} = \int_{E_1}^{E_2} \sigma (h\nu) dh\nu
\end{equation}
is the resonance strength.  Evaluating the continuum oscillator strength 
$f$ for the ALS cross section measurements yielded  a value of 5.743 $\pm$ 1.150,
assuming an error of 20\%. 
 A similar procedure for the theoretical Dirac {\it R}-matrix cross sections gave a
value of 4.966 for 98\% of the $3d^{10}4s\; {\rm ^2S}_{1/2}$ ground state
and 2\% of the statistical average of the  $3d^{9}4s^2\; {\rm ^2D}_{3/2}$ 
and $3d^{9}4s^2\; {\rm ^2D}_{5/2}$ metastable states, 
in good agreement with experiment.  We note that for an  individual 
resonance strength $\overline{\sigma}_{\rm PI}$,  the energy grid  [$E_1$, $E_2$] 
is over only the energy span of the resonance. We have used this procedure to 
determine the resonance strengths from our work. 

For the peaks that are well resolved, in the \citet{Peart1987} data,  
resonance strengths have been derived and are compared 
with values obtained from the ALS measurements and the DARC calculations.
The results are tabulated in Table \ref{tab2}.  
In general, the resonance strengths of \citet{Peart1987} are much larger than 
the corresponding values determined from our work. It is seen that only two resonant 
peaks (with comparable values) are within the uncertainties of 
both measurements, i.e., the two first peaks of group $n$=5.

It is important to point out that \citet{Peart1987} used an 
electron bombardment ion source of zinc vapor to generate their Zn$^{+}$ ion beam. 
In a study of emission cross sections for electron-impact excitation of Zn$^{+}$, \citet{Rogers1982} used the 
same type of hot-filament ion source with vaporized zinc metal. They found a large background signal due to 
higher-lying metastable Zn$^{+}$ ions in their beam. Metastable contamination 
in the Zn$^{+}$ ion beam was lowered by reducing the ion-source anode voltage to 12 V.   
\citet{Peart1987} studied the effect on the population of electronic excitation in their ion 
source by changing the voltage of the  anode of their ion source. 
In this type of  ion source, the amount of electronic 
excitation can be controlled by changing the  anode voltage, 
since at low pressures, this voltage corresponds 
to the electron impact kinetic energy.   In studies
 of emission cross sections for electron impact excitation of Zn$^{+}$ ions, \citet{Rogers1982} 
showed a difference of less than 20\% in the excitation cross section between the 
voltages that \citet{Peart1987} used in their comparison (13.4 V and 23.0 V). 

This relatively small change in the electron-impact excitation cross section of Zn$^{+}$ for the two electron 
kinetic energies, combined with a substantial population of ground state Zn$^{+}$, implied that the ion beam 
was in almost the same internal energy at both anode voltages. 
This explains why they did not observe any measurable differences in 
their photoionization cross section as a function of the anode potential.  
Electron impact generates electronic excitation.  A possible explanation 
for the disagreement in the resonance strengths is that the ion beam population of electronically 
excited states of \citet{Peart1987} were different than the population of 
excited states in the ion beam of the present ALS experiment.

\section{Conclusions}                                                                                                                                                                    
Photoionization cross sections producing Zn$^{2+}$  
from Zn$^{+}$ ions were measured in the photon energy range of 
17.5 eV -- 90 eV with a photon energy resolution of 17 meV FWHM.  
All the sharp resonance features present in the spectrum (for the energy region 
20 -- 28 eV) have been analyzed and identified.  These resonance features we 
 attribute to Rydberg transitions originating from the ground state Zn$^{+}$ into 
excited autoionizing states of  the Zn$^{+}$ ion that ionize 
to the different $3d^94s\; {\rm ^3D}_{3,2,1}$ and $3d^94s\; {\rm ^1D}_2$ 
threshold states of Zn$^{2+}$. We note that the DARC calculations, for resonance 
energies and peak heights,  favour the present ALS measurements as opposed 
to the previous measurements from the Daresbury radiation facility \citep{Peart1987}.

The most prominent peak in the spectrum is observed at 24.141 eV in the ALS data, 
24.172 eV in the Daresbury data \citep{Peart1987} and 24.158 eV in the DARC calculations, 
a difference of 31 meV and 17 meV respectively, see Table \ref{tab2}.
We attribute this peak to an excited Zn$^{+}[3d^9 4s({\rm ^3D}_3)\;5p]$ autoionizing resonant state, 
originating  from photoexcitation of the $3d^{10}4s\; {\rm ^2S}_{1/2}$ ground state and
decaying back to the state $3d^94s\; ({\rm ^3D}_3)$ of Zn$^{2+}$.  This autoionizing state
is a member of the Rydberg resonance series I, see panel (a) of Fig. \ref{fig2}.    
We note that transitions from the electronically excited 
$3d^{10}4p\; {\rm ^2P^{o}}_{1/2,3/2}$ levels are dipole allowed to the 
$3d^{10}4s\; {\rm ^2S}_{1/2}$ ground state and will  not be present in the spectrum.

The cross section below the ground state threshold at 17.964 eV does not go to zero, due 
to a small fraction of excited metastable states present in the Zn$^{+}$  ion beam.   
In addition, contributions from higher-order radiation components may contaminate the photon beam. 
A comparison of the present results with the earlier measurements of \citet{Peart1987} reproduces 
all their resonant peaks but with much improved statistics.  Discrepancies in the cross sections 
in the low-energy regime of the spectrum we attribute to the different 
populations of the excited and metastable states in the ion beams of both experiments.   

Interpretation of the present data was  made possible using photoionization cross section results  determined 
from large-scale Dirac $R$-matrix calculations which allows for 
the inclusion of the interaction  between closed and open channels. 
We traced the contributions from different initial states in the ion beam and determined 
the metastable fraction present in the experiment to be small, in the region of 2\%.
\section*{Acknowledgements}
The Advanced Light Source is supported by the Director, 
Office of Science, Office of Basic Energy Sciences, of the 
U. S. Department of Energy under Contract No.DE-AC02-05CH11231.  
AMC acknowledges support through Co-operative Agreement DOE-NA0002075. 
GH acknowledges Reyes Garc\'{i}a and Ulises Amaya for computational support, 
grant PAPIIT-IN-109-317 and DGAPA-PASPA sabbatical scholarship. 
DC acknowledges support from Sierra College and 
OW from the Swedish Research Council.  
BMMcL acknowledges support by the US National Science Foundation 
under the visitors program through a grant to 
ITAMP at the Harvard-Smithsonian Center for Astrophysics 
and Queen's University Belfast through a visiting research fellowship (VRF).
This research used resources of the National Energy Research Scientific Computing Centre, 
which is supported by the Office of Science of the U.S. Department  of Energy  (DOE) 
under Contract No. DE-AC02-05CH11231. 
The computational work was performed at the National Energy Research Scientific
Computing Centre in Berkeley, CA, USA and at The High Performance 
Computing Centre Stuttgart (HLRS) 
of the University of Stuttgart, Stuttgart, Germany. 
Grants of computing time at NERSC and HLRS 
are gratefully acknowledged.

%
\bibliographystyle{mnras}                       
\bibliography{znplus}
%
\bsp	
\label{lastpage}
\end{document}